\documentclass[prc,preprint]{revtex4}
\usepackage{epsfig}
\usepackage{graphics}
\usepackage{amsmath}
\tighten

\renewcommand{\vec}[1]{\mathbf{#1}}

\begin{document}
\bibliographystyle{revtex}

\title{Orbital magnetism in axially deformed sodium clusters:
From scissors mode to dia-para magnetic anisotropy}
\author{
V.O. Nesterenko$^{1}$, W. Kleinig$^{1,2}$, P.-G. Reinhard$^{3}$,
N. Lo Iudice$^{4}$, F.F. de Souza Cruz$^{5}$, and
J.R. Marinelli$^{5}$
}

\affiliation{
$^{1}$
     Bogoliubov Laboratory of Theoretical Physics,
     Joint Institute for Nuclear Research,
     Dubna, Moscow region, 141980, Russia
\\$^{2}$
     Technische Universit\"at Dresden,
     Institut f\"ur Analysis, Dresden, D-01062, Germany
\\$^{3}$
     Institut f\"ur Theoretische Physik,
     Universit\"at Erlangen, D-91058, Erlangen, Germany
\\$^{4}$
     Dipartimento di Scienze Fisiche,
     Universit${\grave{a}}$ di Napoli ''Federico II'' and Istituto
     Nazionale di Fisica Nucleare, Dipartamento di Scienze Fisiche, Monte S.
     Angelo, Via Cinthia I-80126, Napoli, Italy
\\$^{5}$
     Departamento de Fisica, Universidade Federal de Santa
     Catarina, Florian$\acute o$polis, SC, 88040-900, Brasil
}

\date{\today}
\begin{abstract}
Low-energy orbital magnetic dipole excitations, known as scissors mode
(SM), are studied in alkali metal clusters.  Subsequent
dynamic and static effects are explored.  The treatment is based on
a self-consistent microscopic approach using the jellium
approximation for the ionic background and the Kohn-Sham mean field
for the electrons. The microscopic origin of SM and its main features
(structure of the mode in light and medium clusters, separation into
low- and high-energy plasmons, coupling high-energy M1 scissors and E2
quadrupole plasmons, contributions of shape isomers, etc) are
discussed. The scissors M1 strength acquires large values with
increasing cluster size. The mode is responsible for the van Vleck
paramagnetism of spin-saturated clusters. Quantum shell effects induce
a fragile interplay between Langevin diamagnetism and van Vleck
paramagnetism and lead to a remarkable dia-para anisotropy in magnetic
susceptibility of particular light clusters. Finally, several routes
for observing the SM experimentally are discussed.
\end{abstract}

\maketitle

\section{Introduction}

This paper deals with orbital magnetism in metal clusters.  Because
of a possibly large number of atoms, the valence electrons may accede
single-particle orbitals with very high angular momenta. The
occupation of these orbitals has a large impact in cluster static
magnetism \cite{LS_ZPD,SRL,magrev} or in collective magnetic modes of
orbital nature \cite{LS_ZPD,Ne_sc,Re_M1,Ne_twist}. Two
remarkable examples are the scissors \cite{LS_ZPD,Ne_sc,Re_M1} and
twist \cite{Ne_twist} modes.  The scissors mode is
strictly correlated with cluster deformation.  It can be viewed as a
small-amplitude rotational oscillation of a spheroid of valence
electrons against a spheroid of the ionic background (hence the name SM).

The SM is a general dynamical phenomenon already found or predicted in
different quantum finite systems.  It was first proposed \cite{Iu_M1}
and observed \cite{richter} in atomic nuclei where it still remains a
hot topic for both experimental and theoretical studies (for a review
see \cite{I97}). It was later predicted in a variety of different
systems, like metal clusters \cite{LS_ZPD,Ne_sc}, quantum dots
\cite{QD_M1} and ultra-cold superfluid gas of fermionic atoms
\cite{FA_M1}. More remarkably, it was predicted \cite{BE_M1} and
observed \cite{BEexp_M1} in a Bose-Einstein condensate.
All these different systems have two features in common: The broken
spherical symmetry and the two-component nature (neutrons and protons
in nuclei, valence electrons and ions in atomic clusters, electrons
and surrounding media in quantum dots, atoms and the trap in dilute
Fermi gas and Bose condensate).

The SM strongly affects orbital magnetism in alkali metal clusters.
It can be described in terms of
Larmor diamagnetism and temperature-independent van Vleck
paramagnetism \cite{LS_ZPD,SRL}. They are both weak and, therefore,
need to be studied in systems, like alkali metal clusters, where
strong forms of magnetism, like ferromagnetism, are absent. Being a
low-energy mode, the SM determines the van Vleck paramagnetism and
causes a strong anisotropy in the magnetic susceptibility
\cite{LS_ZPD,Ne_Rich}. Moreover, some particular light clusters can
exhibit dia-para anisotropy, being diamagnetic along the z symmetry
axis and paramagnetic in x-, y-directions \cite{Ne_Rich}.

The SM has already been studied in schematic \cite{LS_ZPD} and
microscopic \cite{Ne_sc,Ne_Rich,Ne_SN} approaches. The microscopic
calculations, though accounting for quantum shells effects, were not
fully self-consistent. A deformed Woods-Saxon \cite{Ne_Tsu}, rather
than a self-consistent Kohn-Sham, one-body potential was adopted. The
quadrupole deformation was deduced from other models or experimental
estimates.
Certainly, we need to perform fully self-consistent
calculations based on density-functional theory in order to settle
the subtle issues, like the fragile dia-para anisotropy, the role of
the ionic structure, triaxiality, shape isomers, etc..

Self-consistent calculations, accounting for the ionic structure, were
performed for spin-saturated ground and spin-polarized isomeric states
of light sodium clusters ${{\rm Na}_{12}}$ and ${{\rm Na}_{16}}$
\cite{Re_M1}. It was found that the scissors response remains
determined basically by the global deformation, in spite of the fact
that triaxiality and ionic structure induce a strong fragmentation in
the strength. It was also shown that the detailed ionic structure
destroys locally spherical symmetry thereby causing a finite, though
very weak, M1 response (transverse optical mode) even in clusters with
zero global deformation.

Calculations of the same level of completeness become prohibitive as
one moves to heavier clusters.  For this reason, we are forced to use
in the present paper the Kohn-Sham approach with a soft jellium model
for the ionic background \cite{ja}.  This simplifies greatly the
calculations and allows to proceed to heavier clusters. At the same
time, the jellium approach is accurate enough for the principle
problems considered here.  The treatment of the electrons is fully
self-consistent. We adopt a deformed Kohn-Sham mean field using the
energy functional of \cite{Gun}. The cluster shape (in terms of axial
quadrupole and hexadecapole deformations) is determined by varying the
jellium deformation and minimizing the total energy of the system.
The optical response in the linear regime is calculated within the
separable random-phase-approximation (SRPA) method \cite{Ne_PRA,Ne_AP}
self-consistently derived from the Kohn-Sham functional. The SRPA has
been already successfully employed for the description of the dipole
plasmon in spherical \cite{Ne_EPJ_98} and deformed
\cite{Ne_AP,Ne_EPJ_02} alkali metal clusters.

In the present paper, we will consider both the optical M1 response
and static magnetic orbital effects. In Section 2, macroscopic and
microscopic treatments of the SM are briefly outlined. In Section 3
the calculation scheme is presented. The optical response is discussed
in Sections 4 and 5. It will be shown that, in analogy with the
electric dipole plasmon in deformed clusters \cite{Ne_AP,Ne_EPJ_02},
the M1 response in light clusters has a distinctive profile determined
by the deformation. Instead, due to a strong Landau damping and
contributions of shape isomers, the response in medium clusters
becomes vague and the SM can be viewed as a statistical mix of
contributions from different cluster shapes.  Besides, we discuss
structure of the residual interaction and explain a small collective
shift in SM excitations.  The coupling between the high-energy SM
branch and the electric quadrupole plasmon is demonstrated. In Section
6, the dia-para anisotropy is discussed. In Section 7, we estimate
perspectives to observe the SM in photo-absorption, Raman scattering
and inelastic electron scattering.  The conclusions are given in
Section 8.

\section{The scissors mode: Brief outline}
\label{sec:sciss_brief}

The macroscopic and microscopic treatment of the SM in clusters is
discussed in detail in Refs. \cite{LS_ZPD,Ne_sc,Re_M1}. Thus, we give
here only a brief outline for a better understanding of the results
presented in the next sections.

In the geometrical model of \cite{Iu_M1}, the SM arises from a
rotational oscillations of valence electrons versus the ions, both
assumed to form distinct spheroids (see left part of
Fig. 1).
Following the alternative view of \cite{LS_ZPD}
(right part of Fig. 1, the displacement field
of the mode is a sum  of the rigid rotation and a quadrupole term
(the latter provides vanishing velocity of electrons at the surface):
\begin{equation}
{\vec u}({\vec r})={\vec \Omega}\times {\vec r}+ \delta_2
(1+\delta_2/3)^{-1}%
{\bf \nabla}(yz)
\label{zisp}
\end{equation}
where $\delta_{2}$ is the
quadrupole deformation parameter (to be defined in the next section).
In axially symmetric systems, the SM is generated by the orbital
momentum fields $L_x$ and $L_y$ perpendicular to the symmetry axis
$z$, and it is characterized by the quantum numbers $|\Lambda^{\pi
}=1^{+}>$ where $\Lambda$ is the eigenvalue of $L_z$ and $\pi$ is the
space parity. Energy and magnetic strength of the mode can be
estimated macroscopically \cite{LS_ZPD,Ne_sc}:
\begin{equation}
\omega = \frac{20.7}{r_{s}^{2}}N_{e}^{-1/3}\delta_{2} \ eV,
\label{eq:om}
\end{equation}
\begin{eqnarray}
B(M1) &=&4\langle 1^{+}\mid {\hat L}_{x}\mid 0\rangle^2\mu
_{b}^{2}
\nonumber\\
&=&\frac{2}{3}N_e<r^2> \omega \mu _{b}^{2}
\\
 & \simeq& N_{e}^{4/3}\delta_{2} \ \mu _{b}^{2}
\nonumber
\label{eq:B(M1)}
\end{eqnarray}
where $N_e$ is the number of valence electrons
and $r_s$ the Wigner-Seitz radius. We use here natural units
$m_e=\hbar =c=1$. The value $B(M1)$ stands for summed strength of the
degenerated x- and y-branches. The z-branch vanishes for symmetry
reasons. It is worth noting that $B(M1)$ does not depend on $r_s$ and
so is the same for different metals.

The microscopic treatment of the SM yields the shell structure of an
axially deformed mean field. One can characterize the emerging
single-electron states in terms of the quantum numbers of the axially
deformed harmonic oscillator (Clemenger-Nilsson basis). These are the
triplets $\nu=[{\cal N} n_z \Lambda ]$ where $n_z$ labels the number of
nodes in $z$-direction (=symmetry axis)
and ${\cal N}$ is the
principle shell number ${\cal N}=n_z+2n_r+\Lambda$ (from which one can
derive the number $n_r$ of radial nodes).
The angular momenta orthogonal to the symmetry axis, ${\hat L}_{x}$
and ${\hat L}_{y}$, promote low-energy $\Delta {\cal N}=0$ transitions
inside the valence shell and high-energy $\Delta {\cal N}=2$ transitions
across two shells. Moreover, one may expand wave functions in terms of
the spherical  basis  $(n L\Lambda )$
\begin{equation}
 \Psi_{\nu =[{\cal N}n_z\Lambda ]}
 =
 \sum_{nL} a^{\nu}_{nL}R_{nL}(r) Y_{L\Lambda}(\Omega).
\end{equation}
This allows to evaluate the single-particle orbital M1 transition
amplitude between hole ($\nu =h$) and particle ($\nu =p$) states
\begin{eqnarray}
 && \langle\Psi_{p}|{\hat L}_{x}|\Psi_{h}\rangle \propto
\delta^{\mbox{}}_{\pi_{p},\pi_{h}}\delta^{\mbox{}}_{\Lambda_{p},
\Lambda_{h}\!\pm\!1}
\label{eq:me}\\
 &&
\qquad\qquad
\sum_{nL}
a^{p}_{nL}a^{h}_{nL}\sqrt{L(L\!+\!1)\!-\!\Lambda_h(\Lambda_h\!\pm \!1)}.
 \nonumber
\end{eqnarray}
Eq. (\ref{eq:me}) shows that the scissors mode is generated by
$\Lambda_p=\Lambda_h\pm 1$ transitions between the components of one
and the same spherical $(nL)$-level. In spherical systems,
$(nL\Lambda)$-states belonging to the level $(nL)$ are degenerate
while in deformed systems they exhibit the deformation splitting and
so may be connected by M1 transitions with non-zero excitation
energies. This is the origin of the scissors mode.  The energy scale
of the scissors mode is determined by the deformation energy splitting
and so is rather small. This explains the predominantly low-energy
($\Delta \cal{N}$=0) character of the scissors mode. Just the
low-energy branch carries most of the scissors $B(M1)$ strength
(see also discussion in Section \ref{sec:scissresp}). The high-energy
($\Delta \cal{N}$=2) branch of the mode is much weaker since the
particle states involved into ($\Delta \cal{N}$=2) transitions include
only small $(nL)$-components from the valence shell.

\section{Calculation scheme}

Our approach \cite{Ne_AP} employs the Kohn-Sham equations for the
electronic mean field using actually the energy-density functional of
\cite{Gun}. The positive ionic background is modeled by a soft jellium
density
\begin{equation}
\label{eq:densi}
 \rho_I({\bf r})=
 \frac{\rho_{I0}}{1+exp((r-R(\theta ))/\alpha )}
\end{equation}
where quadrupole and hexadecapole axial deformations are introduced through
the jellium radius as
\begin{equation}
\label{eq:rad}
 R(\theta )=R_0 \left[1+
\delta_{2} Y_{20}(\theta )+
\delta_{4} Y_{40}(\theta ) \right] .
\end{equation}
The optimal deformation parameters $\delta_{2}$ and $\delta_{4}$ are
determined by minimizing the total energy.

\begin{table}
\caption{
\label{tab:deform}
Ground state deformation parameters $\delta_2$
and $\delta_4$ and moments $\beta_2$  and $\beta_4$.
For ${\rm Na}_{55}^+$ and ${\rm Na}_{119}^+$ the
deformation parameters for the isomeric states
together with  their energy deficits $\Delta E$
are also  given.
}
\centering
\begin{tabular}{|c|c|c|c|c|c|}\hline
Cluster & $\delta_2$ & $\delta_4$ & $\beta_2$  & $\beta_4$ & $\Delta E$, eV
\\ \hline
 ${\rm Na}_{11}^+$  & 0.355  & 0.25  & 0.44  & 0.41 & - \\
${\rm Na}_{15}^+$  & 0.59  & -0.19  & 0.47  & -0.02 & - \\
${\rm Na}_{19}^+$  & -0.285  & -0.09  & -0.21  & -0.02 & - \\
 ${\rm Na}_{27}^+$  & 0.33  & 0.08 & 0.36  & 0.17 & - \\
 ${\rm Na}_{35}^+$  & -0.21 & 0.02 &-0.18  & 0.04& - \\
${\rm Na}_{55}^+$  & 0.18  & -0.07  & 0.17  & -0.05 & - \\
              & -0.11 & -0.07 & -0.09 & -0.05 & 0.020 \\
${\rm Na}_{119}^+$  & -0.27  & -0.14  & -0.20  & -0.06 & - \\
              & 0.24 & -0.04 & 0.24 & $\sim$ & 0.004 \\
            & -0.04 & -0.22 & -0.02 & -0.18 & 0.024 \\
\hline
\end{tabular}
\end{table}

We consider the clusters ${\rm Na}_{11}^+$, ${\rm Na}_{15}^+$,
${\rm Na}_{19}^+$, ${\rm Na}_{27}^+$, ${\rm Na}_{35}^+$, ${\rm
Na}_{55}^+$, and ${\rm Na}_{119}^+$, which, according to jellium
estimates \cite{Ne_AP,Ne_EPJ_02,Rein_def,Yan_def,Thom_def,Pash_def},
exhibit axial deformations. These clusters represent a wide size region
and, as shown in Table I,
cover prolate (${\rm
Na}_{11}^+$, ${\rm Na}_{15}^+$, ${\rm Na}_{27}^+$, ${\rm
Na}_{55}^+$) and oblate (${\rm Na}_{19}^+$, ${\rm Na}_{35}^+$,
${\rm Na}_{119}^+$) shapes in ground states. Moreover, few of them
(${\rm Na}_{55}^+$ and ${\rm Na}_{119}^+$) have shape isomers with
a tiny energy deficit $\Delta E \sim 0.02$ eV \cite{Ne_AP,Ne_EPJ_02}
and with quadrupole deformation of opposite sign with respect to the
ground state. The largest sample ${\rm Na}_{119}^+$ has also a
hexadecapole isomer.

Table I
also shows the multipole moments
\begin{eqnarray}
\beta_{\lambda}
 &=&
\frac{4\pi}{3}
\frac{\int d{\bf r}\rho_0({\bf r}) r^{\lambda}Y_{\lambda 0}}{N_e {\tilde R}^\lambda}
\label{eq:QL}
\;,\quad
{\tilde R}
=
\sqrt{\frac{5}{3} \frac{\int d{\bf r}\rho_0({\bf r}) r^2 }
{\int d{\bf r}\rho_0({\bf r})}}
\end{eqnarray}
where $\lambda =2,4$ and $\rho_0({\bf r})$ is the ground state density
of valence electrons.  The dimensionless multipole moments
$\beta_{\lambda}$ are less model dependent because they are computed
from expectation values. Thus they can serve for robust
characterization of the deformation and for comparison between
different models. The quantities $\delta_{\lambda}$ in the jellium
model (\ref{eq:densi})  coincide with the
$\beta_{\lambda}$ for small deformation.

The optical response is calculated in the framework of the random
phase approximation (RPA). Full RPA for deformed systems is extremely
involved. We employ a separable approximation of RPA where the residual
interaction is expanded into a sum of separable terms \cite{Ne_AP}.
The expansion employs local one-body operators $Q_{\lambda 1p}(\vec{r})$.
Their structure is constructed as to map the response mean-field in
RPA and the generating operators
(to which response is explored)
are chosen to cover the leading multipole operators.  The expansion
coefficients are computed self-consistently.  It was shown that this
procedure provides a sufficiently precise reproduction of the exact
residual interaction \cite{Ne_AP}.
In particular, for the description of the SM, we use the basis of
generating operators
\begin{eqnarray}
f_{21p}&=&r^{2+p}(Y_{21}(\theta )+Y_{21}^{\dagger}(\theta )), \qquad
p=0,2,4,
\nonumber\\
\label{eq:iop}
f_{41p}&=&r^{4+p}(Y_{41}(\theta )+Y_{41}^{\dagger}(\theta )), \qquad
p=0,2.
\end{eqnarray}
The same set of operators was used for the description of
$\lambda\mu=21$ branch of the quadrupole plasmon \cite{Ne_AP}.  The
close similarity between scissors and quadrupole fields was discussed
in \cite{Ne_sc}. The $p=0$ component of the input field $f_{210}$ has
a form of the external quadrupole field in the long-wave
approximation.  It generates the main piece of the separable
interaction, peaked at the surface of the system. The next two
quadrupole fields ($p=2$ and 4) result in the separable
operators $Q_{21p}(\vec{r})$ peaked in more interior of the
cluster. We include also hexadecapole fields $f_{41p}$  in order to
account for
coupling between quadrupole and hexadecapole modes, arising with the
onset of deformation, especially in systems having both quadrupole and
hexadecapole deformations. The set of the fields (\ref{eq:iop})
ensures good convergence of the separable expansion to exact results.
Explicit expressions for $Q_{\lambda 1p}(\vec{r})$ are given
in \cite{Ne_AP}.

We study the SM response in terms of photo-absorption.
In axially deformed systems, the photo-absorption cross-section
from the ground state to the excited state  $j=\Lambda^{\pi}$  of
excitation energy $\omega_j$ is
\begin{equation}
\label{eq:M1_cs}
\sigma (X\lambda\mu, gs \!\rightarrow\! j)=
\frac{8\pi^3\lambda\!+\!1}{\lambda[(2\lambda\!+\!1)!!]^2}
(\frac{\omega_j}{\hbar c})^{2\lambda -1}
|<j|\hat{O}^X_{\lambda\mu}|gs>|^2
\end{equation}
where $<j|\hat{O}^X_{\lambda\mu}|gs>$ is the reduced transition matrix
element and $\hat{O}^X_{\lambda\mu}$ is the operator of electric
$(X=E)$ or magnetic $(X=M)$ transition. For the scissors mode, we have
$\hat{O}^M_{11}=\hat{l}_x$ (in Bohr magnetons $\mu_B$).  The
selections rules are $\mu =\Lambda$ and $(-1)^{\lambda}=\pi$ for
$(X=E)$ or $(-1)^{\lambda +1}=\pi$ for $(X=M)$.  The electric
photo-absorption strength (\ref{eq:M1_cs}) will be used in Section
\ref{sec:exper} for estimation of the competition between the scissors
and low-energy electric excitations.

Useful measures for the SM are provided by the sum rules
\begin{equation}
\label{eq:sr}
S_m(M1)=\sum_j \omega^m_j B(M1)_j
\end{equation}
for $m=-1,0,$ and 1.  The ratios $\omega =\sqrt{S_{1}/S_{-1}}$ or
$\omega = S_{1}/S_{0}$ provide a rough energy centroid of the
low-energy mode. The $S_{-1}$ is proportional
to the paramagnetic susceptibility
and $S_1$ to the integral photo-absorption cross section.
The detailed values for the $S_m$ are obtained from explicit
RPA results. A simple
estimate for $S_1$ can be obtained using Eqs. (\ref{eq:om}) and
(\ref{eq:B(M1)}):
\begin{equation}
\label{eq:s1}
   S_1(M1)=\sum_j \omega_j B(M1)_j=
   \frac{20.7}{r_s^2} N_e \delta^2_2 \mu_b^2 .
\end{equation}

Sum rules do exist also for electric excitations
where they can be reduced to the simple expressions \cite{Ne_PRA}
\begin{eqnarray}
\label{eq:esr}
S_1(E\lambda) &=&
\sum_j \omega_j  |<j|er^{\lambda}Y_{\lambda\mu}|gs>|^2 \\
    &=&
\frac{\hbar^2e^2}{8\pi m_e}\lambda (2\lambda +1)^2 N_e <r^{2\lambda -2}> .
\nonumber
\end{eqnarray}

\section{Scissors response}
\label{sec:scissresp}

The M1 optical responses in light clusters are shown in Fig. 2.
The present Kohn-Sham calculations yield results close to the ones
obtained by using a deformed Woods Saxon potential \cite{Ne_sc}. Only
the excitation energies in prolate clusters are $\sim 0.2$ eV higher,
because of the larger quadrupole deformations obtained in the
self-consistent approach.
The main characteristic of the M1 response is the occurrence of one or
two prominent peaks below 1 eV. They are only slightly shifted from
their unperturbed $1ph$ (particle-hole) spectrum. In the low-energy
region, the 1ph $|\Lambda^{\pi }=1^{+}>$ spectrum is very dilute and,
therefore, does not meet the conditions for Landau fragmentation or
for pronounced coherent superpositions.
Because of these features, the low-energy scissors strength can be
associated to well defined $1ph$ transitions. In the Nilsson-Clemenger
notation $[{\cal N} n_z \Lambda ]$, they are $ [110] \rightarrow
[101]$ in ${\rm Na}_{11}^+$, $[211] \rightarrow [202]$ and
$[211] \rightarrow [200]$ in ${\rm Na}_{15}^+$, $[211]
\rightarrow [220]$ in ${\rm Na}_{19}^+$, $[321] \rightarrow
[310]$ and $[321] \rightarrow [312]$ in ${\rm Na}_{27}^+$,
$[310] \rightarrow [321]$ and $[312] \rightarrow [321]$ in
${\rm Na}_{35}^+$.

As  shown in Fig. 2 (compare full and dashed lines), the
residual interaction induces a rather moderate blue-shift that is much
smaller than e.g. the shift for the electric dipole plasmon.  The
underlying physics is explained in Fig. 3 where the
exchange-correlations and Coulomb contributions to the leading part of
the separable operator $Q_{211}(\vec{r})$ are presented for the case
of ${\rm Na}_{19}^+$ (the repulsive Coulomb is negative in this
representation). It is seen that both contributions compensate
each other to a large extent. The final outcome is a slight repulsive
interaction responsible for the blueshift. It is worth noting that the
balance between these two contributions is rather fragile and can be
affected by different factors, e.g. triaxiality and detailed ionic
structure, making even the sign of the net interaction uncertain
\cite{Re_M1}.

\begin{table}
\caption{
Sum rules $S_{m}$ (in $eV^m \mu _{b}^{2}$) calculated in the energy
region 0-6 eV. The fractions for the region 0-1 eV are given
in parenthesis.
}
\label{tab:sum_rules}
\centering
\begin{tabular}{|l|l|l|l|}\hline
 Cluster      & $S_{-1}$  & $S_{0}$   & $S_{1}$  \\ \hline
 ${\rm Na}_{11}^+$  & 16.5 (92$\%$) & 14.7 (72$\%$) & 21.5 (34$\%$)\\
 ${\rm Na}_{15}^+$  & 18.7 (84$\%$) & 19.2 (69$\%$)   & 25.9 (43$\%$)\\
 ${\rm Na}_{19}^+$  & 11.3 (98$\%$) &  8.2 (90$\%$)   & 7.4 (67$\%$)\\
 ${\rm Na}_{27}^+$  & 80.6 (96$\%$) & 35.6 (74$\%$)   & 37.7 (25$\%$) \\
 ${\rm Na}_{35}^+$  & 38.9 (97$\%$) & 17.2 (84$\%$)   & 13.3 (43$\%$) \\
 ${\rm Na}_{55}^+$  & 94.9 (97$\%$) & 32.6 (78$\%$)   & 30.2 (26$\%$) \\
 ${\rm Na}_{119}^+$ & 544 (97$\%$) & 138 (80$\%$)     & 103 (36$\%$) \\
\hline
\end{tabular}
\end{table}

Table II
collects summed scissors strengths in
wide- (0-6 eV) and low-energy (0-1 eV) regions, calculated within
the SRPA. It is seen that the $\Delta {\cal N}=0$
low-energy scissors mode, being mainly concentrated below 1 eV,
contributes strongly to $S_{-1}$ and  $S_{0}$. This justifies using
 $S_{-1}$ and  $S_{0}$ for a rough estimation of the energy of
the mode as $\omega =\sqrt{S_{1}/S_{-1}}$ (or $\omega = S_{1}/S_{0}$).
Besides, since  $S_{-1}$ is proportional to the
paramagnetic susceptibility, this means that just the low-energy
SM determines the van Vleck paramagnetism (see discussion in
Section \ref{sec:mag_an}).
The high-energy part of the scissors strength (associated with
$\Delta {\cal N}=2$ transitions)
contributes appreciably to the $S_{1}$ sum rule, i.e. to the total
photo-absorption cross section. Table II
also shows
that the increase of the $S_m$ with cluster size is not monotonous.
The fluctuations are caused by the predominantly  1ph character of the
low-energy SM and by the difference in cluster deformations. ${\rm Na}^+_{27}$
demonstrates especially strong M1 strength, while in oblate clusters
${\rm Na}^+_{19}$ and ${\rm Na}^+_{35}$  the strength is relatively weak.

Fig. 4 illustrates the coupling of the high-energy
scissors branch to the quadrupole plasmon in ${\rm Na}_{55}^+$. The
correlation between M1 and E2 peaks at $\sim 3$ eV is clearly
seen. The coupling of electric and magnetic modes with the same
quantum numbers $\Lambda^{\pi }$ is a general feature of deformed
finite quantum systems. It is well known, for example, in atomic
nuclei (see, e.g. \cite{I97,Kva_PRC}).

The coupling of the SM with dipole and spin-dipole oscillations in
clusters was discussed in detail in Ref. \cite{Re_M1}.  In particular,
it was shown that breaking the symmetry by the ionic lattice results
in a weak coupling between these oscillations of opposite space
parity. This feature may change for clusters deposited on a surface
since symmetry breaking is much stronger there.

\section{Effects of shape isomers and ionic structure}
\label{sec:isomers}

The calculations \cite{Ne_AP,Ne_EPJ_02} show that light axially
deformed clusters ($N_e < 40$) have at the deformation-energy surface
one distinct minimum corresponding to the ground-state
deformation. The heavier clusters ($40 < N_e < 100$ ) have usually two
minima with opposite quadrupole deformations (prolate and oblate) and
very close energies.  The energy difference between the ground and the
shape isomeric states is often less than 0.02 eV = 200 K. The number
of shape isomers with a tiny energy deficit increases with cluster
size.  These isomers may have a variety of very different shapes.
And significant amounts of isomers can be found in a thermal ensemblse
and thus contribute to the SM, e.g., at room temperature.

Fig. 5 compares the scissors modes built on
the ground and isomeric states in ${{\rm Na}_{55}}^+$. Both of them
show  a rich
spectrum of low lying M1 states. But the spectra and typical strengths
of the two states differ substantially. Indeed, the M1 spectra are
generated by several $1ph$ M1 transitions between the levels lying in
the vicinity of the Fermi surface.  Prolate and oblate shapes yield
different sequences of the single-particle levels just at the
Fermi surface. As a result, some of the $1ph$ M1 transitions
which are significant in the prolate case
are transformed to $1pp$ or $1hh$ transitions in the oblate case,
thus strongly decreasing the strength.

Fig. 5 compares also the SRPA results obtained
with soft jellium model (\ref{eq:densi}) and detailed
ionic structure \cite{Re_M1}. In the case of the ionic structure,
the global cluster deformations slightly deviate from those
in the jellium case. It is seen that calculations with the ionic structure
give somewhat different positions of the SM peaks with respect to the
jellium case.  Such a redistribution of the peaks is due to different
energies of the single-particle levels in the ionic calculations.
The photo-absorption response in the ionic case is stronger, which is also
explained by the redistribution of the levels in the mean field. The level
sequence in this case favors a few more $1ph$ transitions of M1 type.

Altogether, the calculations demonstrate some important points concerning
the SM in medium clusters:\\[4pt]
i) The mode is distributed over several $1ph$ transitions. \\[4pt]
ii) The single particle spectrum  near the Fermi level is rather dense
and so even small changes in the calculation scheme can redistribute
visibly the spectrum and open (or close) some relevant $1ph$ transitions.
\\[4pt]
iii)
As was mentioned above, the ground and first isomeric states in
medium clusters can be very close in energy.
 So, the scissors mode in free medium clusters should be considered
as a statistical mixture of contributions from different shapes
(predominantly of the ground state and first isomer).
Such kind of the analysis
(involving the deformation splitting,
Landau fragmentation and contributions of shape isomers) was recently
applied to explain experimental E1 optical response in deformed sodium
clusters with $50<N_e<60$ \cite{Ne_EPJ_02}.\\[4pt]
Less ambiguous deformations will be provided by the SM in metal
clusters deposited on insulating substrates \cite{Tra}. These clusters
are oblate and their mean size and magnitude of the deformation can be
well controlled. Such clusters seem to be the most
promising systems for experimental search of the SM.

\section{Magnetic anisotropy}\label{sec:mag_an}

RPA calculations show that the SM energies and B(M1) strengths scale
with the deformation $\delta_2$ and the electron number $N_{\rm e}$ basically
according to the trends (\ref{eq:om}) and (\ref{eq:B(M1)}).  Strong
fluctuations, however, take place in small clusters \cite{Ne_sc}. They
reflect the $1ph$ nature of the transitions and may affect the magnetic
susceptibility.

The total orbital magnetic susceptibility in clusters is the sum
of Langevin diamagnetic and van Vleck paramagnetic terms
\cite{LS_ZPD,SRL}:
\begin{equation}
\chi_k =\chi^{dia}_k + \chi^{para}_k,
\label{ms}
\end{equation}
where
\begin{equation}
\chi^{dia}_k =- \mu_b^2 N_e <\rho^2_k> = -\mu_b^2
\Theta_k^R,
\end{equation}
\begin{equation}
\label{eq:chi_para}
\chi^{para}_k=2\mu_b^2\sum_j \frac{|<j|{\hat L}_k|0>|^2}{\omega_j}
= \mu_b^2 \Theta_k,
\end{equation}
having denoted by
\begin{equation}
\label{eq:mom_in}
\Theta_{k}=2\sum_j \frac{|<j|{\hat L}_{k}|0>|^2}{\omega_j}
\end{equation}
the moment of inertia and by
\begin{equation}
\Theta_{k}^R =N_e <\rho^2_{k}>
\end{equation}
its rigid-body value. The sum in (\ref{eq:chi_para}) and
(\ref{eq:mom_in}) runs over excited states. Further, $k = x,y,z$ is
the coordinate index. In axial systems, one has $\rho^2_z=2< x^2>$ and
$\rho^2_{x,y}=<x^2>+<z^2>$.

Note that for $k=x,y$ the operator entering in
the matrix element in (\ref{eq:chi_para}) is
exactly the scissors generator. This makes evident that just the low-energy
SM mainly contributes to $\chi^{para}_{x,y}$. Indeed Table II
shows that contribution of the low-energy scissors mode to the value
$S_{-1} \sim \chi^{para}_k$ achieves $85 -100\%$. So, just the SM
determines the van Vleck paramagnetism.

In the schematic model
\cite{LS_ZPD}, the moment of inertia comes out as the rigid-body
value, so that $\theta_{x,y}=\theta_{x,y}^R$ and, therefore,
$\chi^{para}_{x,y}= - \chi^{dia}_{x,y}$, i.e. a complete
compensation of dia- and para-magnetic terms in $\chi_{x,y}$ takes
place.  Due to axial symmetry, one also has $\chi^{para}_z=0$.  The
total susceptibility becomes, therefore, strictly anisotropic
\cite{LS_ZPD}
\begin{equation}\label{eq:bal}
\chi_x=\chi_y=0, \qquad \chi_z=\chi^{dia}_z,
\end{equation}
going from zero to diamagnetic values.

On the other hand, strong shell effects in some peculiar light
clusters may alter appreciably the above result.  This is illustrated
for ${\rm Na}_{27}^+$ in Fig. 6. Because of very
low excitation energy of the SM in this cluster (see
an exceptionally large value
of $S_{-1}$ for ${\rm Na}_{27}^+$ in Fig. 2
and, also, in Table II,
the paramagnetic susceptibility
is enhanced considerably and is no longer balanced by the diamagnetic
term. So, ${\rm Na}_{27}^+$ should be paramagnetic in x,y-directions
and diamagnetic in z-direction.  The cluster ${\rm Na}_{11}^+$ also
hints this property. The magnetic moments in ${\rm Na}_{27}^+$ are
sufficiently large to allow for a measurement of the dia-para
anisotropy. The observation of this effect would provide a strong
(though indirect) evidence of the SM in clusters.

\section{Experimental perspectives}
\label{sec:exper}

\subsection{General analysis}

As was mentioned in the Introduction, the SM is not yet observed
experimentally in metal clusters. The search of the SM is hindered by
several factors:

1) The mode has very low photo-absorption cross section.
Following our results, in sodium clusters with $N_e\simeq 10 - 10^2$ ,
$\sigma (M1)/N_e\simeq 10^{-5}-10^{-7} \AA^2$
as compared to $\sigma (E1)/N_e\simeq 2\AA^2$ for the dipole plasmon.
Such a weak M1 signal is at the edge of the sensitivity of modern
detectors.

2) Eqs. (\ref{eq:M1_cs})-(\ref{eq:esr}) allow to estimate
the ratios for maximal optical responses as
\begin{equation}
\label{eq:e1m1}
\sigma (E1)/\sigma (M1)=0.35 \cdot 10^5 \AA^{-2}(\frac{r_s}{\delta_2})^2
\end{equation}
and (for $\omega_{E2}\simeq 3$ eV \cite{Ne_PRA,Ne_AP})
\begin{equation}
\label{eq:e2m1}
\sigma (E2)/\sigma (M1)=
  0.82 \cdot 10^{-2} \AA^{-4} (\frac{r_s^2}{\delta_2})^2 N_e^{2/3} .
\end{equation}
This gives for deformed ($\delta_2 =0.2$) sodium ($r_s=2.1 \AA$)
clusters
$\sigma
(E1)/\sigma (M1)\simeq 4 \cdot 10^6$ and $\sigma (E2)/\sigma
(M1)\simeq 10^2 $
(the latter for $N_e =125$).
So, the SM suffers from the competition with E1 and E2 strengths.
Besides, the competition with E2 increases with cluster size. The
above estimates compare the SM optical response with the maximal
responses of E1 and E2 plasmons. As is shown in the next subsection,
the competition is much weaker in the low-energy region where the SM
has its stronghold, but one has still to be aware of large amounts of
E1 strength.

3) The low-energy scissors mode lies in the infrared region where
commonly used detectors for a visible light are not efficient enough.

4) The SM energy decreases with cluster size and reaches the region of
phonon excitations, $\sim 0.1 eV$,  in very big clusters.

In spite of these hindrances, clusters offer enough opportunities to
look for the optimal conditions for observing the SM: one may change
cluster size and (or) deformation, use different metals, choose
between free and supported clusters, etc..  The macroscopic estimates
(\ref{eq:om})-(\ref{eq:B(M1)}) and the present jellium RPA
calculations may serve as a guide.  Such analysis is briefly given
below.  To this purpose, we consider, as a typical case, sodium
clusters with a moderate deformation $\delta_2 =0.2$.

Larger clusters help because the M1 photo-absorption cross section
grows linearly with the size.  Already in deformed clusters with
$N_e=10^4-10^6$ the scissors signal should be detectable.
Unfortunately, the corresponding energy $\omega \simeq 0.1-0.01$ eV
approaches or even covers the region of phonon excitations.
Thus the best compromise for free sodium clusters is achieved at $N_e
\sim 10^3$. Besides that, these clusters are large enough to ensure
the dominance of the orbital scissors mode over spin M1 excitations.

We may also look for clusters of larger density and deformation to
increase the energy and the strength of the mode. For example, Li
clusters ($r_s=1.7$ \AA) allow to increase both the energy and optical
response by a factor of $\sim 1.5$ with respect to Na ($r_s=2.1$
\AA). Further improvement may be achieved with Ag ($r_s=1.6$ \AA), Mg
($r_s=1.4$ \AA), or Al ($r_s=1.1$ \AA). In any case, highly deformed
clusters are welcome because $\sigma (M1)\sim \delta^2_2$ and $\omega
\sim \delta_2$.

In general, free clusters seem not suitable for observing the SM.
They are well mass-separated only up to sizes of hundred
atoms. However, these systems have a weak M1 signal. As for heavier
clusters, they suffer from a poor mass separation and are expected to
be weakly deformed \cite{Pash_def}. Moreover, as was discussed in
Section \ref{sec:isomers}, their M1 signal is a statistical mix of
contributions of different shapes given by the ground state and
isomers.

Deposited clusters look more promising.
One can adopt techniques that allow to get oblate clusters (Na and Ag)
on dielectric surfaces \cite{Tra} and, more remarkably, to monitor
their size and deformation. Clusters with $10^2-10^6$ atoms can be
used for this aim. In this way, one can obtain supported clusters of a
desired size and shape. High density of clusters on the surface gives
good statistics in the measurements. Last but not least, monitoring
the shape of clusters gives the chance to use the trends $\sigma
(M1)\sim \delta_2^2$ and $\omega_{M1} \sim \delta_2$ to distinguish
the SM from the E1 and E2 signals. To this end, one
has to irradiate clusters of the same size but of different
deformations. Then the low-energy E1 and E2 cross sections are also
affected but, hopefully, in an irregular fashion (thus averaging
out) or, at least, in a way different from the trends mentioned above.
The cross sections measured at different deformations
can be  mutually subtracted to extract a small useful signal.
For the analysis of such experiments, we need accurate theoretical
estimates accounting for all the main factors and, in particular,
the influence of the surface on the electric and scissors plasmons.
Such work is in progress.

The competition with E1 and E2 modes may be bypassed partly by
resorting to specific reactions and techniques capable of hindering if
not suppressing the E1 and E2 channels.  We will discuss this issue
for three relevant reactions: photo-absorption, resonance Raman
scattering, and inelastic electron scattering.

\subsection{Photo-absorption}

The SM lies much lower than the dipole and quadrupole plasmon modes
(0.2-1.0 eV against 2.5-3.0 eV for the dipole and 2-4 eV for the
quadrupole \cite{Ne_PRA,Ne_AP}). In principle, one could exploit this
energy separation to focus on the SM. However, its M1 photo-absorption
cross section is still very small as compared to the E1 strength and
can be masked by the tail of close E1 modes.
This is illustrated in Fig. 7 which compares E1, E2
and M1 photo-absorption strengths calculated within SRPA
approach\cite{Ne_AP} in axially deformed ${\rm Na}_{119}^+$
(ground state). The E1
strength, though extremely small compared to the dipole plasmon peak,
is still strong enough to mask the SM.  The competition with the
quadrupole E2 strength in ${\rm Na}_{119}^+$ is negligible (though
it can be stronger in heavier clusters).

The competition with the E1 channel can be decreased by using deformed
(oblate) clusters with $N\simeq 10^3-10^4$ atoms, supported on
dielectric surfaces, and adopting infrared techniques with polarized
light. For the SM, the magnetic field of the incoming light should be
parallel to the surface. This can be done with both s- and p-polarizations
(with the electrical field to be  perpendicular and parallel
to the incident plane, respectively). The s-polarized light under an
angle $\sim 0^{\circ}$ with the surface normal  suppresses the $\mu =0$
E1 branch without weakening the SM.  The p-polarized light at angle
$\sim 90^{\circ}$ suppresses the  $\mu =1$ E1 branch, again without
weakening the SM. The second variant seems to be preferable
since the $\mu =1$ branch is generally stronger
(see, e.g., Fig. 7.

\subsection{Resonant Raman scattering}

In resonance Raman scattering (inelastic scattering of polarized
photons), the SM may be populated through the de-excitation of the
dipole plasmon excited by the incoming photon. The difference between
energies of incoming and outgoing photons gives the energy of the
mode.  One must look at E1 decays, which represent the leading channel
and may populate in deformed systems M1 as well as E2 states.
With respect to photo-absorption, resonance Raman scattering has the
advantage of excluding the competition with the strong E1 mode and to
work in the visible light region.  An experiment of this kind was
carried out in supported Ag clusters \cite{Duval}.  The analysis of
the data suggested, however,  that the E2 rather than the SM was
populated through the E1 decay.  On the other hand, it is not simple
to separate clearly
the E2 from the M1 modes which are mixed in deformed systems.

\subsection{Inelastic electron scattering}

Inelastic electron scattering is also a promising technique. At
backward scattering angles, the M1 signal should prevail over the electric
ones. This property enabled the discovery of the scissors mode in
deformed atomic nuclei \cite{richter}. Polarization of the electrons
provides additional possibilities to extract the magnetic response.
In nuclear physics, electrons with
very high energies (tens and hundreds MeV) are used, which favors
creation of well collimated, intense and monochromatic electron
beams. Moreover, at so high energies, it is easier to enhance the
relative contribution of the transverse form factor.

The low-energy, intense and  monochromatic electron beams,
suitable for our purpose, as well as the angular resolved
techniques are now available \cite{Birm}. To our knowledge, however,
inelastic electron scattering was used so far only to observe the
dipole plasmon. Theoretical estimates \cite{Sol} show that at certain
scattering angles the dipole contribution can be well suppressed in
favor of the quadrupole plasmon, which, therefore, can be observed.
From the latter analysis, we may infer that also the SM may be
extracted. Indeed, in the low-energy region the SM
can be stronger excited in the photo-absorption than the E2 mode.
The analogy with atomic
nuclei suggests that it should be favored even more in electron
scattering at backward angles.

\section{Conclusions}

The scissors mode (SM) in axially deformed sodium clusters has been
studied within a self-consistent RPA approach based on the soft
jellium model for the ionic background and density-functional theory
for the electrons. The calculations show that the distribution of
the low-energy scissors strength is dominated by $1ph$ states with only
small shifts by the residual interaction. The SM thus reflects almost
directly the density of $1ph$ M1 transitions near the Fermi surface and
exhibits an extreme sensitivity to the actual deformation.
The SM is rather unambiguous in light clusters ($N_e<40$) while
it should be a statistical mix of contributions from different shapes
in heavier clusters. Furthermore, the high-energy branch of the SM in
deformed systems is coupled with the quadrupole plasmon.

The calculations show that the SM determines the Van Vleck
paramagnetism and results in a strong anisotropy of the magnetic
susceptibility. Moreover, due to quantum shell effects, some clusters
(${\rm Na}_{27}^+$) demonstrate dia-para anisotropy when the axial
cluster is diamagnetic along the symmetry axis and paramagnetic along
other two axes. The magnetic moments are large enough to be
measured. The experimental observation of the dia-para anisotropy,
being interesting itself, could also serve as an indirect evidence of
the SM in clusters.

A general analysis of possible routes for experimental observation of
the scissors mode was done. The photo-absorption with a polarized
light, inelastic electron scattering and Raman scattering were
considered.  The former two reactions seem to be more promising.
Moreover, deposited clusters favor the detection of the SM. Its
features need yet to be worked out in detail.  Work along that
direction is in progress.

{\bf Acknowledgement:}\\
The work was partly supported (V.O.N.) by RFBR (00-02-17194), Heisenberg -
Landau (Germany-BLTP JINR) and DFG (436RUS17/102/01) grants.

\newpage

{\bf \large FIGURE CAPTIONS}

\vspace{0.5cm}\indent
{\bf Figure 1}:
Macroscopic view of scissors mode : rigid rotation \cite{Iu_M1} (left),
and rotation within a rigid surface \cite{LS_ZPD} (right).

\vspace{0.5cm}\indent
{\bf Figure 2}:
Photo-absorption cross-section for the scissors mode, weighted by the
Lorentz function with the averaging parameter 0.25 eV. The responses
with (solid curve) and without (dashed curve) the residual interaction
are presented.

\vspace{0.5cm}\indent
{\bf Figure 3}:
Radial profile of the residual interaction in ${\rm Na}_{19}^+$.
Exchange-correlation (long-dashed curve) and Coulomb
(shot-dashed curve) contributions are given together with
their sum (solid curve).

\vspace{0.5cm}\indent
{\bf Figure 4}:
Coupling between the high-energy scissors branch and
E2 quadrupole plasmon in prolate ${\rm Na}_{55}^+$.
M1 and E2 photo-absorptions are weighted by the
Lorentz function with the averaging parameter 0.25 eV.

\vspace{0.5cm}\indent
{\bf Figure 5}:
The strength distribution of the scissors modes in ${\rm Na}_{55}^+$,
built on the prolate ground state (upper panel) and oblate
first isomer (lower panel). Jellium (solid curve) and ionic
(dashed curve) SRPA results are compared.

\vspace{0.5cm}\indent
{\bf Figure 6}:
Normalized diamagnetic, paramagnetic and summed moments
$\mu =\chi B$ ($B=4 T$) in axial deformed clusters
${\rm Na}_{11}^+$, ${\rm Na}_{15}^+$, ${\rm Na}_{19}^+$,
${\rm Na}_{27}^+$, ${\rm Na}_{35}^+$,
${\rm Na}_{55}^+$ and ${\rm Na}_{119}^+$.

\vspace{0.5cm}\indent
{\bf Figure 7}:
Photo-absorption E1, E2 and scissors cross sections
in ${\rm Na}_{119}^+$ weighted by the
Lorentz function with the averaging parameter 0.25 eV.
The E1 $\mu =0$ and 1 branches are given separately
to show their relative contributions to the competition
with the scissors mode.

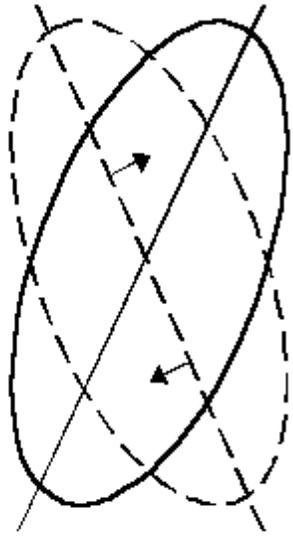
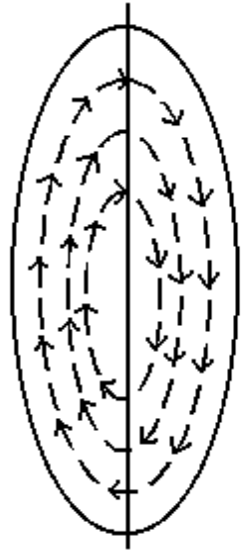
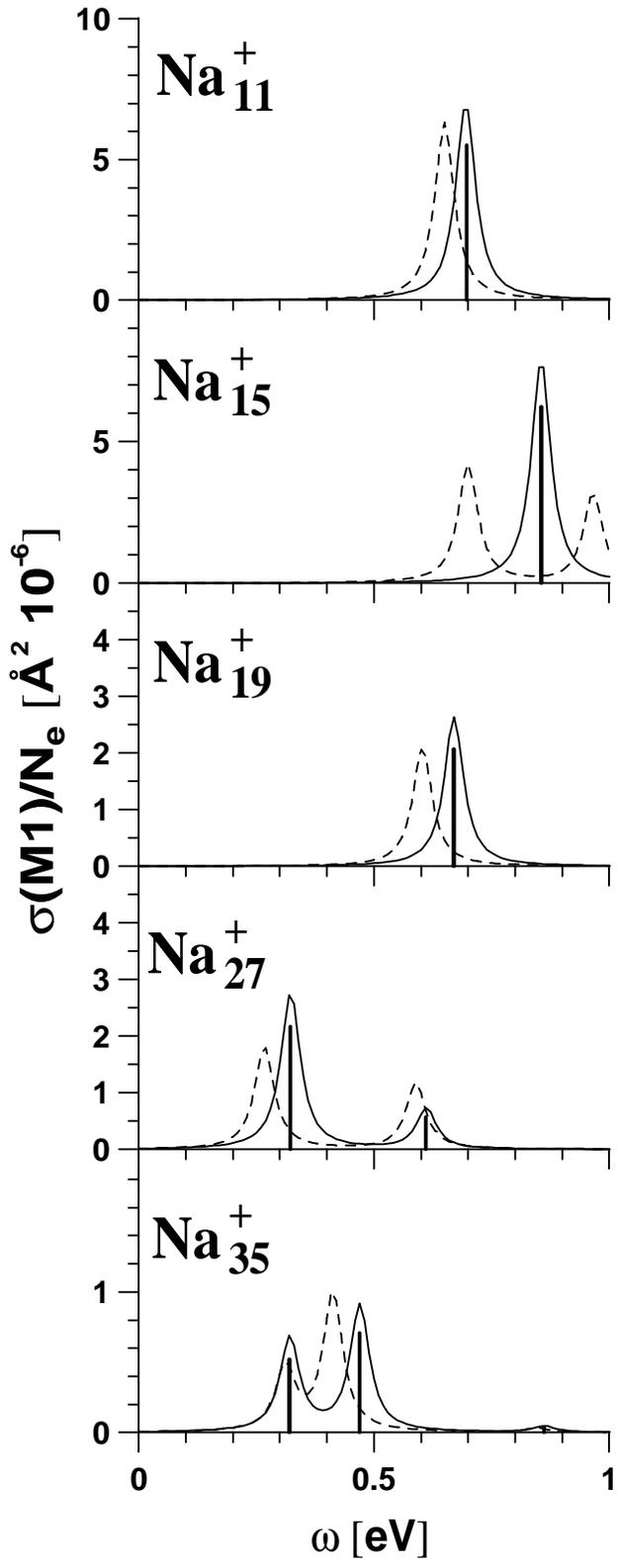
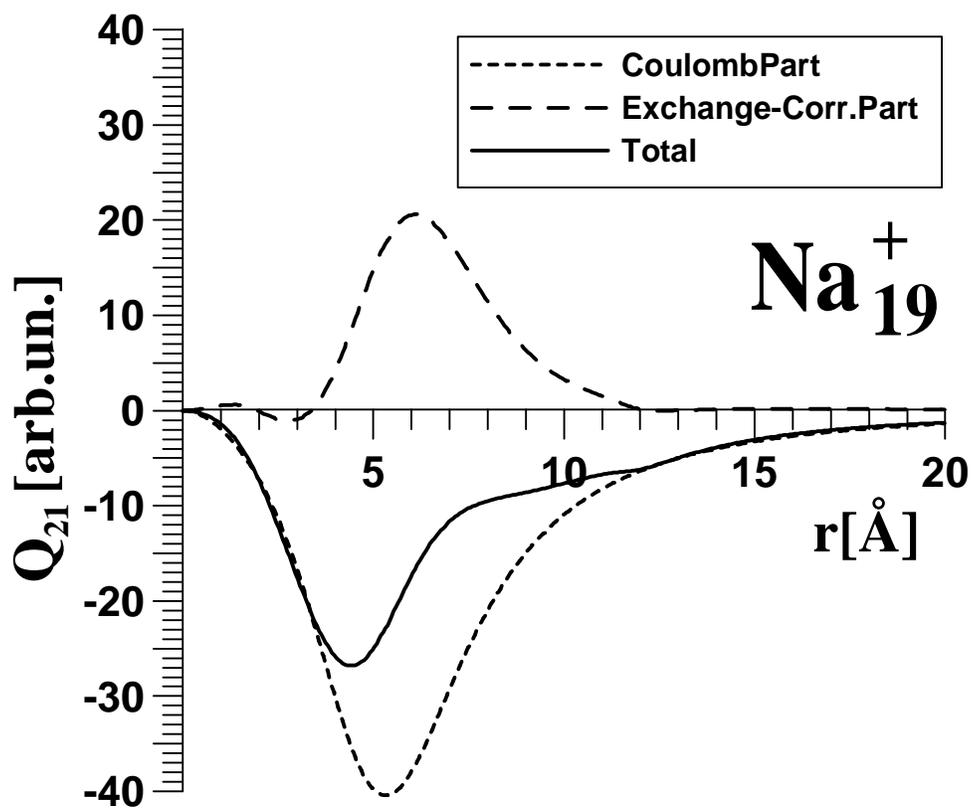
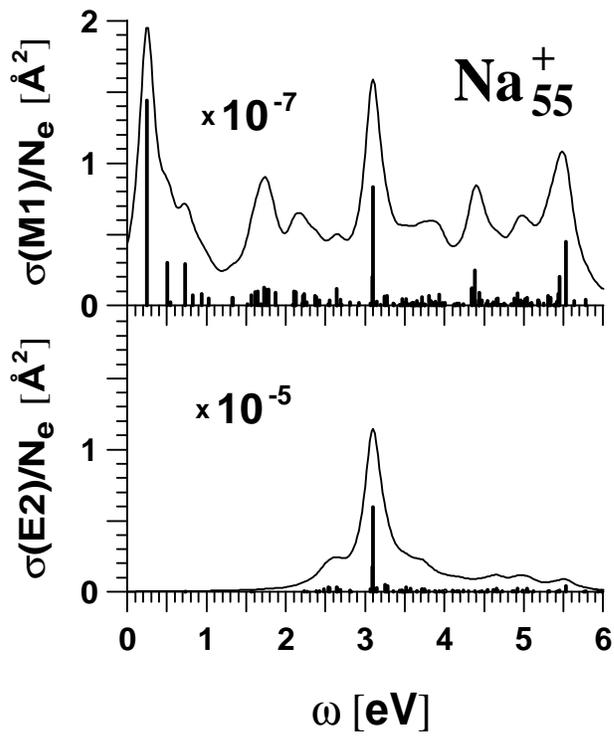
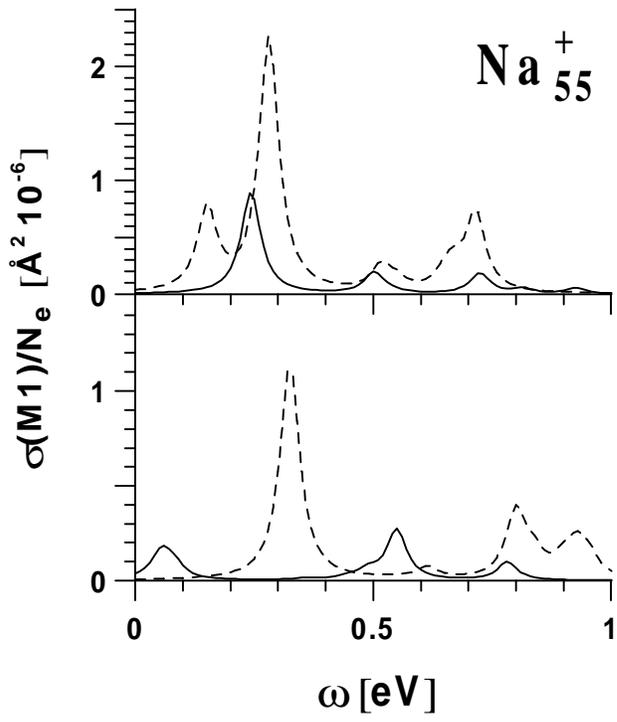
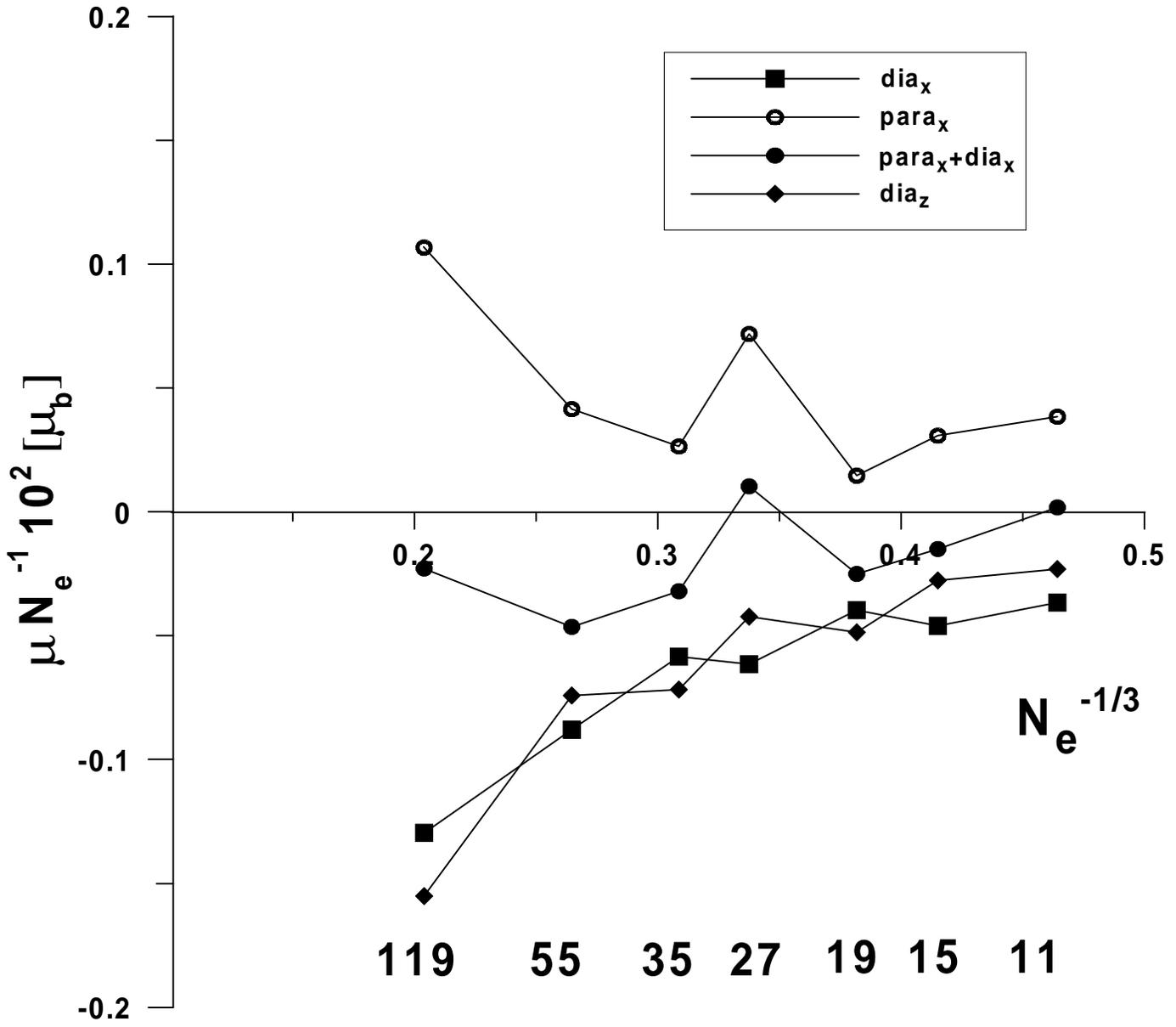
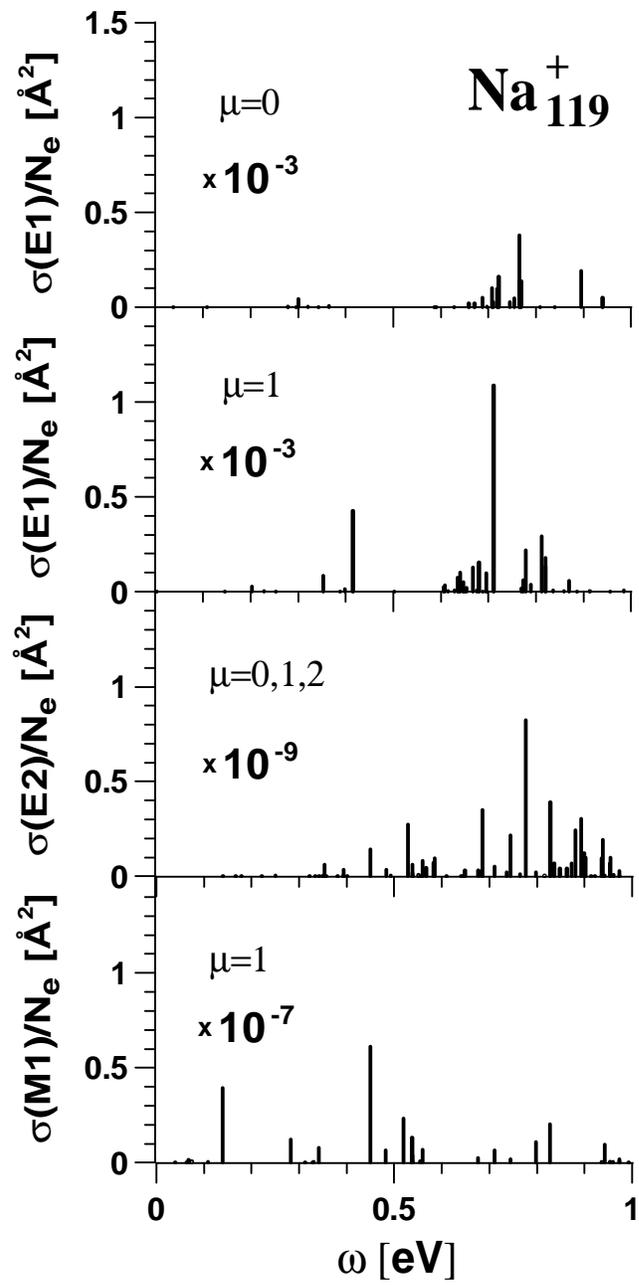
\end{document}